# A Multiport Approach to Thermal Noise and Scattering Parameter Simulation of Cryogenic Experiments

Maurio B. Grando, Christian R. Boutan, Jihee Yang

¹ **In this paper, a simple algorithm for detailed system-level thermal noise analysis is developed, demonstrated, and verified. This method uses noise-wave theory and noise covariance matrices to cascade noise and scattering parameters of multiport devices at different temperatures. This method addresses the effects of return loss, multiport isolation/coupling, and static temperature differentials between components, and will work in cases where the noise temperature is at or near the quantum noise limit. An ideal multidevice network will first be demonstrated to show that this method's results are consistent with the Friis cascade when component parameters such as return loss and isolation are ideal. Following the ideal multidevice example, a cryogenic experiment is conducted to demonstrate that the proposed simulation method is successful when real data are used.**

## I. INTRODUCTION

NOISE-wave theory, first introduced in 1981 [1], provides a reliable way to account for thermal noise in both passive [2], [3] and active components [4]. Noise-wave analysis builds upon the more familiar voltage scattering parameters (s-parameters) to describe a noise model for microwave devices, a powerful concept for analyzing noise in linear circuits.

This work describes how multiport noise waves can be easily cascaded, like s-parameters. We present a general case of the embedded element method for noise waves [3], [5] to develop a simple algorithm that cascades multiport noise matrices while simultaneously solving for the system s-parameters. This method is helpful in multiport systems analysis and is particularly useful for complicated networks of passive and active devices in a system whose components span a range of static temperatures. Furthermore, the proposed method also accounts for the quantum-limited noise floor imposed by uncertainty principle.

This high-level algorithm is intended to aid a systems engineer in modeling and analyzing the effects of component return loss and isolation near the thermal noise floor without relying on a Friis cascade approximation [6] which assumes two ports and highly isolated components.

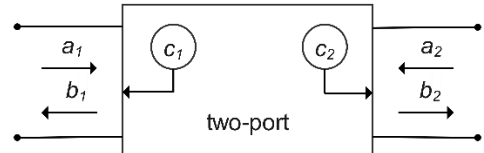

Fig. 1. Noise-wave theory matrix schematic for a generic two-port device. Incoming ($a_1$, $a_2$) and outgoing ($b_1$, $b_2$) waves are shown. Internal noise sources ($c_1$, $c_2$) are shown. Subscripts 1 and 2 denote port numbers.

## II. GENERAL THEORY FOR CASCADING MULTIPORT PARAMETERS

Though the principles are extensible to multiport devices, the classic example of noise-wave theory is demonstrated featuring a two-port device, shown in Fig. 1. While incoming waves ($a_1$, $a_2$) originate from an external stimulus, ($b_1$, $b_2$) are a combination of reflected noise from $a_1$ and $a_2$ and internal noise sources at each port ($c_1$, $c_2$). Noise sources ($c_1$, $c_2$) are used to describe internally generated noise within the device. These noise sources can represent thermal noise as well as other linear noise sources (like shot noise) emanating from within a device. These internal sources, together with interaction from external noise sources as described by s-parameters, fully describe the outgoing and incoming noise waves.

$$\begin{bmatrix} b_1 \\ b_2 \end{bmatrix} = \begin{bmatrix} S_{11} & S_{12} \\ S_{21} & S_{22} \end{bmatrix} \begin{bmatrix} a_1 \\ a_2 \end{bmatrix} + \begin{bmatrix} c_1 \\ c_2 \end{bmatrix} \quad (1)$$

where the $S_{ij}$ are the s-parameters. Or in a more abstract form,

$$\boldsymbol{b} = \boldsymbol{S}\boldsymbol{a} + \boldsymbol{c}. \quad (2)$$

When the noise sources $c_1$ and $c_2$ are not independent of each Invictus Animus Research and Design, LLC, Centennial, CO 80122 USA (e-mail: maurio.grando@ianimus.net).

C. R. Boutan is with Pacific Northwest National Laboratory, Richland, WA 99354 USA (e-mail: christian.boutan@pnnl.gov).

J. Yang is with Pacific Northwest National Laboratory, Richland, WA 99354 USA (e-mail: jihee.yang@pnnl.gov).

other, the noise covariance matrix is needed. Therefore, it is advantageous to define the noise covariance matrix $C$. This matrix describes the correlation between the noise source at each port relative to every other port. The two-port example of the noise covariance matrix $C$ is shown in (3).

$$C = \overline{cc^\dagger} = \overline{\begin{bmatrix} c_1 \\ c_2 \end{bmatrix} \begin{bmatrix} c_1^\dagger & c_2^\dagger \end{bmatrix}} = \begin{bmatrix} \overline{|c_1|^2} & \overline{c_1 c_2^\dagger} \\ \overline{c_2 c_1^\dagger} & \overline{|c_2|^2} \end{bmatrix} \quad (3)$$

where $c$ is the noise source vector $\begin{bmatrix} c_1 \\ c_2 \end{bmatrix}$ from (1), $\dagger$ is the Hermitian conjugate operator, and the bar indicates the time average. Diagonal elements in the noise covariance matrix $C$—later shown to be particularly useful for calculating the noise temperature and noise figure—represent the noise (over a 1 Hz bandwidth) that is present when the respective port is terminated by a perfectly matched and noiseless termination. The off-diagonal pairs provide information about the correlation between noise sources at each port. The noise covariance matrix is typically symmetric meaning that repeated information is carried in half the off-diagonal entries (conjugate of each other). An off-diagonal entry of zero indicates no correlation between the two ports. The following sections will build upon this theory and the noise covariance matrix $C$.

Here, we extend this theory to encompass a multiport solution. To do this, we use the solution to an embedded multiport problem [2], [4] as shown in Fig. 2. The embedded multiport network approach assumes that the scattering matrix $S^{sys}$ and noise covariance matrix $C^{sys}$ are ordered into external and internal submatrices. The dimension of $K$, the embedded scattering matrix, is the same as $S^{sys}$. The matrix $K$ is ordered—internal and external ports—identically to $S^{sys}$. $K$ is an ordered version of the connection matrix described in [5].

$K_{ii}$ is the internal-to-internal submatrix of matrix $K$. In every case only the submatrix $K_{ii}$ will be populated. The other submatrices of $K$ ($K_{ee}$, $K_{ie}$, and $K_{ei}$) are null (zero) matrices for the purpose of this algorithm.

The $S^{sys}$ and $C^{sys}$ are shown in detail below where the subscript "e" indicates an external connection and the subscript "i" denotes an internal connection within the system network. Although they are comparable to the two-port example in Section I, these matrices have been reordered, as described here.

$$b = S^{sys}a + c \quad (4)$$

$$\begin{bmatrix} b_e \\ b_i \end{bmatrix} = \begin{bmatrix} S_{ee}^{sys} & S_{ei}^{sys} \\ S_{ie}^{sys} & S_{ii}^{sys} \end{bmatrix} \begin{bmatrix} a_e \\ a_i \end{bmatrix} + \begin{bmatrix} c_e \\ c_i \end{bmatrix} \quad (5)$$

$$C^{sys} = \overline{cc^\dagger} = \begin{bmatrix} \overline{c_e c_e^\dagger} & \overline{c_e c_i^\dagger} \\ \overline{c_i c_e^\dagger} & \overline{c_i c_i^\dagger} \end{bmatrix} \quad (6)$$

The general embedded network problem has a known solution for the net noise covariance matrix $C^{net}$ and a solution for the net s-parameters $S^{net}$ shown here, summarized in [3]

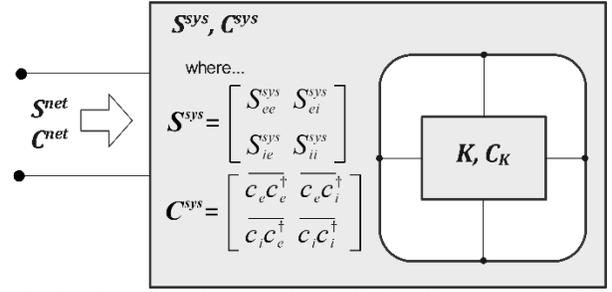

Fig. 2. Illustration of a general embedded multiport network. $S^{sys}$ and $C^{sys}$ represent the scattering and noise covariance matrices for the overall system network. Submatrices of $S^{sys}$ and $C^{sys}$ are shown where the subscript "e" and "i" denote "external" and "internal" connections respectively. An embedded device is represented by scattering and noise covariance matrices $K$ and $C_K$ which only connect to elements within the system network. The entire network is described externally by net scattering and net noise covariance matrices, $S^{net}$ and $C^{net}$.

and [5]:

$$\Lambda = S_{ei}^{sys}(I - K_{ii}S_{ii})^{-1} \quad (7)$$

$$C^{net} = \Lambda C_K \Lambda^\dagger + [I|\Lambda K_{ii}]C^{sys}[I|\Lambda K_{ii}]^\dagger \quad (8)$$

$$S^{net} = S_{ee}^{sys} + \Lambda K_{ii}S_{ie}^{sys} \quad (9)$$

where $I$ is the identity matrix, $C_K$ is the noise covariance matrix of the embedded element, and | is the augmented matrix operator.

Equations (5)–(9) provide a complete solution for the classic embedded network noise analysis problem. We will use this solution as a basis for a method that will allow the analysis of an arbitrary collection of linear multiport devices.

It is beneficial to conceptualize the system multiport network ($S^{sys}$, $C^{sys}$) as a network of arbitrary multiport devices, each with an arbitrary number of ports as depicted in Fig. 3. These arbitrary multiport devices can have connections either outside the embedded multiport or can interface with each other through the embedded scattering and noise covariance matrices ($K_{ii}$, $C_K$). Our algorithm will assume that any port not identified as interfacing to the embedded matrix pair is automatically assumed to be an external port of the net multiport system ($S^{net}$, $C^{net}$).

We then define the embedded device represented by matrices ($K_{ii}$, $C_K$) as a s-parameter and noise network of lossless, passive, and noiseless two-port devices. These devices serve to define lossless, matched connections between ports contained in the multinetwork collection $S^{sys}$ and $C^{sys}$ seen in Fig. 3. Because the embedded network $K_{ii}$ is defined as perfectly matched, it must also have zeros along the diagonal. Knowing that $K_{ii}$ represents a collection of lossless two-port devices, it can be interpreted as a symmetric matrix of only ones and zeros. The ones define where port connections occur between devices in the multidevice network represented by $S^{sys}$ and $C^{sys}$. Defining the embedded network as a noiseless system causes $C_K = 0$. Then we can rewrite (8) as

$$C^{net} = [I|\Lambda K_{ii}]C^{sys}[I|\Lambda K_{ii}]^\dagger. \quad (10)$$

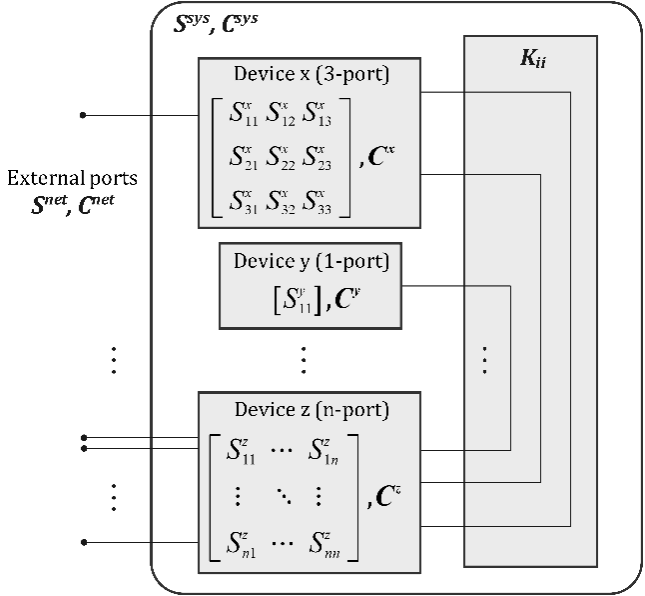

Fig. 3. Re-envisioning the embedded multiport network $K_{ii}$ as a lossless network connecting an arbitrary number of multiport devices represented as a total network $S^{net}$. Here we use letters in the superscript (e.g., $S^x$, $C^x$) to differentiate devices.

It should be noted that eq. (10) is an expanded case of [3], [5] where $C_K$ has been eliminated. This case allows for easy cascading for large systems of sub-matrices.

As a path to a more intuitive understanding of the derivation of $S^{net}$, the expression $\Lambda K_{ii}$ in (9) is equal to the inverse of the connection scattering matrix $W_{ii}$ found in [5] if we abide by our stipulations on $K_{ii}$.

$$W_{ii} = K_{ii} - S_{ii}^{sys} \tag{11}$$

$$S^{net} = S_{ee}^{sys} + S_{ei}^{sys} \, W_{ii}^{-1} S_{ie}^{sys} \tag{12}$$

We now have the foundation for an algorithm that allows the cascading of arbitrary multiport networks of any size using (7), (10), (11), and (12).

## III. NOISE AND S-PARAMETER CASCADE ALGORITHM

To cascade the scattering and noise parameters of multiple devices using this method, you must first assume that each device you plan to add to the multidevice network $S^{sys}$ and $C^{sys}$ has a known scattering matrix (set of s-parameters) and a noise covariance matrix associated with it. Once these parameters are known for each device in your system, the following method will yield the solution for any multiport system consisting of these devices.

### A. List Components in Diagonal Matrices $S^{sys\prime}$ and $C^{sys\prime}$

First, a diagonal matrix is formed with the components in the multiport network. We will call these matrices $S^{sys\prime}$ and $C^{sys\prime}$. The prime denotes that the matrices have not yet been ordered by their external and internal connections. The matrices added to the diagonal of $S^{sys\prime}$ are the individual s-parameters of each device contained in the multiport system. A device can be of any size (number of ports) and these multiport devices can be added to $S^{sys\prime}$ in any order. The same method is used for $C^{sys\prime}$. The order in which devices are added to the diagonals must be the same between $S^{sys\prime}$ and $C^{sys\prime}$. In (13) and (14), the superscript denotes a device, numbered 1–$n$, and the subscript denotes the scattering matrices for arbitrary ports 1–$n$.

$$S^{sys\prime} = \begin{bmatrix} \begin{bmatrix} S^1{}_{11} & \cdots & S^1{}_{1n} \\ \vdots & \ddots & \vdots \\ S^1{}_{n1} & \cdots & S^1{}_{nn} \end{bmatrix} & 0 & 0 \\ 0 & \ddots & 0 \\ 0 & 0 & \begin{bmatrix} S^N{}_{11} & \cdots & S^N{}_{1n} \\ \vdots & \ddots & \vdots \\ S^N{}_{n1} & \cdots & S^N{}_{nn} \end{bmatrix} \end{bmatrix} \tag{13}$$

$$C^{sys\prime} = \begin{bmatrix} \begin{bmatrix} C^1{}_{11} & \cdots & C^1{}_{1n} \\ \vdots & \ddots & \vdots \\ C^1{}_{n1} & \cdots & C^1{}_{nn} \end{bmatrix} & 0 & 0 \\ 0 & \ddots & 0 \\ 0 & 0 & \begin{bmatrix} C^N{}_{11} & \cdots & C^N{}_{1n} \\ \vdots & \ddots & \vdots \\ C^N{}_{n1} & \cdots & C^N{}_{nn} \end{bmatrix} \end{bmatrix} \tag{14}$$

### B. Define Unsorted Connection Matrix $K\prime$

Matrices $S^{sys\prime}$ and $C^{sys\prime}$ cannot be ordered without first defining how the devices are to be connected. The connection matrix $K\prime$ is created to describe the internal connections between devices within the multiport matrix $S^{sys\prime}$. The connection matrix ports (ports 1–$n$) are defined by the order in which individual device s-parameters were entered in the sparse matrix $S^{sys\prime}$ from the previous step in section A. Any port that is not designated as an internal connection in this matrix $K\prime$ will be considered an external port. The diagonals of $K\prime$ must be symmetric, and an entry of one in the matrix indicates a connection between the column port and the row port where that one occurs.

### C. Define Permutation Matrix $P$

Once the connection matrix $K\prime$ is formed, a permutation matrix can be created. This matrix will be used to form $S^{sys\prime}$ and $C^{sys\prime}$ into ordered subarrays $S^{sys}$ and $C^{sys}$. The permutation matrix specifying all internal and external connections can be visualized as selecting rows from the identity matrix (with dimensions equal to the matrices $S^{sys\prime}$ and $C^{sys\prime}$ ) that correspond to the ports where connections are made in the connection matrix $K\prime$ . These rows are then designated as internal connections with subscript "i." The remaining rows of the identity matrix are then automatically designated as external ports with subscript "e." Fig. 4 shows an example of this activity with two two-port devices in the diagonal matrix $S^{sys\prime}$. The permutation matrix $P$ is then used to yield an ordered matrix. $S^{sys}$ for example, as defined in (5), is derived by multiplying matrix $P$ by $S^{sys\prime}$ and the transpose of $P$ as in (15).

$$S^{sys} = P S^{sys\prime} \, P^T = \begin{bmatrix} S_{ee}^{sys} & S_{ei}^{sys} \\ S_{ie}^{sys} & S_{ii}^{sys} \end{bmatrix} \tag{15}$$

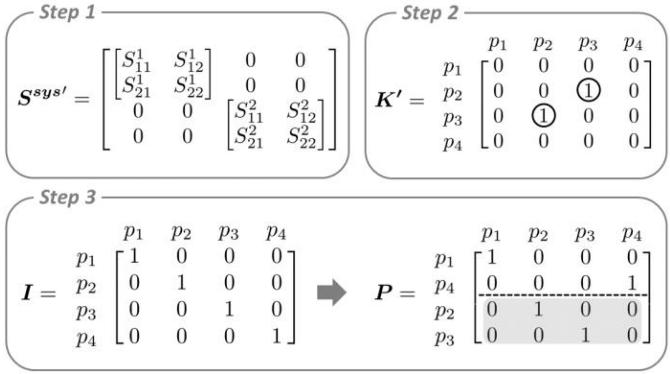

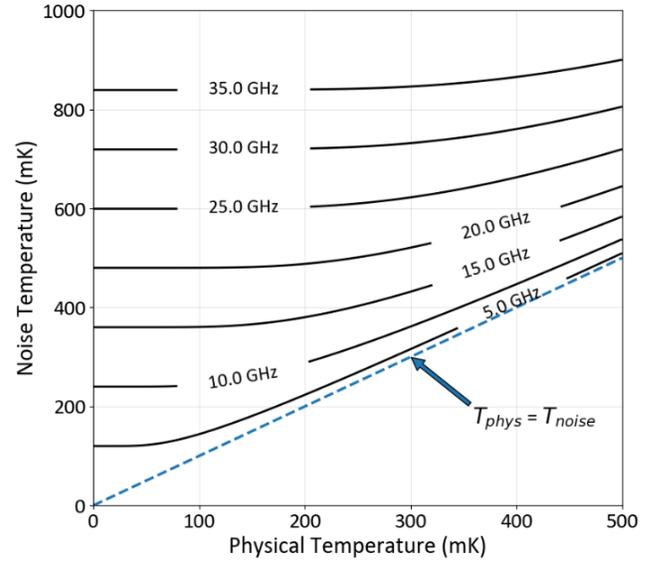

Fig. 4. Step-by-step outline of process to create the permutation matrix. In this example, two (two-port) devices are cascaded together. Port 2 of the first device is connected to port 1 of the second device. The system port numbers are labeled with $\boldsymbol{p_n}$. The dotted line in the permutation matrix $\boldsymbol{P}$ separates between internal and internal ports. The gray box highlights the internal port elements.

The permutation matrix $\boldsymbol{P}$ can be separated into submatrices as well. For the example in Fig. 4, the permutation matrix $\boldsymbol{P}$ can be split into submatrices $\boldsymbol{P_e}$ and $\boldsymbol{P_i}$, as seen in (16) and (17) below. Subscript "e" denotes external ports and "i" denotes internal ports.

$$\boldsymbol{P_i} = \begin{bmatrix} 0 & 1 & 0 & 0 \\ 0 & 0 & 1 & 0 \end{bmatrix} \quad (16)$$

$$\boldsymbol{P_e} = \begin{bmatrix} 1 & 0 & 0 & 0 \\ 0 & 0 & 0 & 1 \end{bmatrix} \quad (17)$$

These submatrices of the permutation matrix can be used to isolate only the submatrix of interest for a given calculation. Use of the permutation matrix is outlined in the next section, where examples are given.

### D. Order Matrices for Calculation

In section C, we defined the permutation submatrices (16) and (17) and their transposes. In this section, we use these permutation submatrices to rearrange the diagonal matrices $\boldsymbol{S^{sys\prime}}$, $\boldsymbol{C^{sys\prime}}$, and $\boldsymbol{K\prime}$ into ordered submatrices needed in (15). Here, all variables needed to calculate $\boldsymbol{S^{net}}$ and $\boldsymbol{C^{net}}$ are derived. Below are the calculations using the permutation matrix to obtain these submatrices called out in (5)–(10).

$$\boldsymbol{S^{sys}_{ee}} = \boldsymbol{P_e}\boldsymbol{S^{sys\prime}}\boldsymbol{P_e^T} \quad (18)$$

$$\boldsymbol{S^{sys}_{ei}} = \boldsymbol{P_e}\boldsymbol{S^{sys\prime}}\boldsymbol{P_i^T} \quad (19)$$

$$\boldsymbol{S^{sys}_{ie}} = \boldsymbol{P_i}\boldsymbol{S^{sys\prime}}\boldsymbol{P_e^T} \quad (20)$$

$$\boldsymbol{S^{sys}_{ii}} = \boldsymbol{P_i}\boldsymbol{S^{sys\prime}}\boldsymbol{P_i^T} \quad (21)$$

$$\boldsymbol{C^{sys}} = \boldsymbol{P}\boldsymbol{C^{sys\prime}}\boldsymbol{P^T} \quad (22)$$

$$\boldsymbol{K_{ii}} = \boldsymbol{P_i}\boldsymbol{K\prime}\boldsymbol{P_i^T} \quad (23)$$

Fig. 5. Noise temperature as a function of physical temperature. Black solid lines represent the noise temperatures from Nyquist theorem of Callen and Welton at different frequencies. The dashed (blue) line represents the noise temperature from the Rayleigh-Jeans approximation ($T_{phys} = T_{noise}$).

### E. Calculate $\boldsymbol{S^{net}}$ and $\boldsymbol{C^{net}}$

At this point, all the requisite variables have been solved for and all that needs to be done is to perform the matrix calculations for the simplified equations derived in (7), (10), (11), and (12). This will yield solutions for $\boldsymbol{S^{net}}$ and $\boldsymbol{C^{net}}$, the system noise covariance and scattering parameters.

## IV. NOISE TEMPERATURE VS. PHYSICAL TEMPERATURE

In room-temperature analysis of thermal noise, it is often acceptable to use a component's physical temperature as the equivalent to noise temperature of a device. This makes the implicit assumption that $hf << k_BT$, or that the energy of a single photon is much less than the surrounding black-body thermal noise. This is known as the Rayleigh-Jeans approximation $P_{noise} = k_BTB$, where $T$ is the physical temperature and $k_B$ is Boltzmann's constant. This is a very good approximation for most frequencies under 1 THz when at room temperature but not necessarily true for other cases. This approximation will not hold, for example, in modern cryogenic systems where the temperature can reach as low as 7–10 mK. At these lower temperatures, the time averaged noise power will plateau at $hf/2k_B$ for a given frequency (see Fig. 5).

As part of this method, any time a noise temperature $T$ is required for noise calculations, we must substitute in the noise temperature defined by the Nyquist theorem of Callen and Welton [7]. This equation is a function of physical temperature and frequency. It will provide an accurate solution for noise temperatures where the Reyleigh-Jean approximation holds as well as conditions where $hf \geq k_BT$ [7], [8].

$$T_n^{CW} = \frac{hf}{2k_B} \coth\left(\frac{hf}{2k_BT}\right) \quad (24)$$

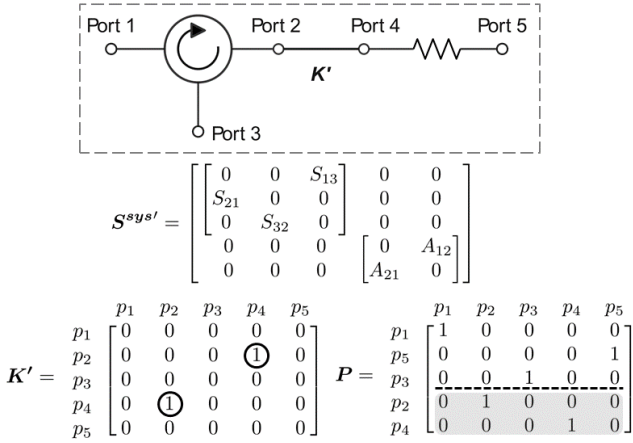

$$S^{sys\prime} = \begin{bmatrix} 0 & 0 & S_{13} & 0 & 0 \\ S_{21} & 0 & 0 & 0 & 0 \\ 0 & S_{32} & 0 & 0 & 0 \\ 0 & 0 & 0 & 0 & A_{12} \\ 0 & 0 & 0 & A_{21} & 0 \end{bmatrix}$$

$$K\prime = \begin{array}{c} \\ p_1 \\ p_2 \\ p_3 \\ p_4 \\ p_5 \end{array} \begin{array}{ccccc} p_1 & p_2 & p_3 & p_4 & p_5 \\ \begin{bmatrix} 0 & 0 & 0 & 0 & 0 \\ 0 & 0 & 0 & 0 & ① \\ 0 & 0 & 0 & 0 & 0 \\ ① & 0 & 0 & 0 & 0 \\ 0 & 0 & 0 & 0 & 0 \end{bmatrix} \end{array} \qquad P = \begin{array}{c} \\ p_1 \\ p_5 \\ p_3 \\ p_2 \\ p_4 \end{array} \begin{array}{ccccc} p_1 & p_2 & p_3 & p_4 & p_5 \\ \begin{bmatrix} 1 & 0 & 0 & 0 & 0 \\ 0 & 0 & 0 & 0 & 1 \\ 0 & 0 & 1 & 0 & 0 \\ 0 & 1 & 0 & 0 & 0 \\ 0 & 0 & 0 & 1 & 0 \end{bmatrix} \end{array}$$

Fig. 6. (Top) Schematic of a multinetwork system with an ideal circulator and attenuator. (Middle, Bottom) Multinetwork matrices, $S^{sys\prime}$, $K\prime$, and $P$, for the circulator-attenuator system. System ports ($p_{1-5}$) have been called out at the rows and columns of matrices. The dotted line in the permutation matrix $P$ separates between external and internal ports. The gray box highlights the internal port elements.

## V. A Simple Example with Ideal Components

In this section we explore a simple multiport problem in detail to give a workable example of the above method. A different physical temperature will be assumed for the two devices to show the effect on the noise temperature equations when devices are at the same temperature and when there is a temperature differential between devices.

Usually, the method outlined in this paper would be programmed so that a user could enter multiple devices into the matrices $S^{sys\prime}$ and $C^{sys\prime}$ and define a connection matrix $K\prime$. The resulting output would be scattering and noise matrices for the system $S^{net}$ and $C^{net}$.

For this example, a multidevice network involving a circulator and attenuator will be considered using the schematic in Fig. 6. We assume the circulator and attenuator are ideal so that the isolation in the circulator is infinite and the ports of both devices are perfectly matched. For an ideal circulator and attenuator, the scattering matrices are

$$S_{circ} = \begin{bmatrix} 0 & 0 & S_{13} \\ S_{21} & 0 & 0 \\ 0 & S_{32} & 0 \end{bmatrix} \quad (25)$$

$$S_{att} = \begin{bmatrix} 0 & A_{12} \\ A_{21} & 0 \end{bmatrix}. \quad (26)$$

Because both devices are passive, Bosma's Theorem [9] can be used to calculate the noise covariance matrix $C$ for each device:

$$C_{circ} = k_B T_{circ}^{cw} \begin{bmatrix} (1-|s_{13}|^2) & 0 & 0 \\ 0 & (1-|s_{21}|^2) & 0 \\ 0 & 0 & (1-|s_{32}|^2) \end{bmatrix} \quad (27)$$

$$C_{att} = k_B T_{att}^{cw} \begin{bmatrix} (1-|A_{12}|^2) & 0 \\ 0 & (1-|A_{21}|^2) \end{bmatrix} \quad (28)$$

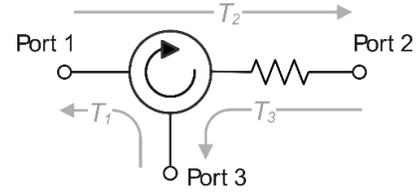

Fig. 7. Schematic of circulator-attenuator multinetwork system (condensed version). Only external ports (1–3) are recognized and shown. Gray arrows represent the direction of transmission for the noise temperature ($T_{1-3}$) calculation at each port.

where noise temperatures of the circulator ($T_{circ}^{cw}$) and attenuator ($T_{atten}^{cw}$), defined by the Nyquist theorem of Callen and Welton and described in (24), need not be the same.

Once each component has an associated scattering matrix and a noise covariance matrix, the sparse diagonal matrix $S^{sys\prime}$ is constructed. Devices can be added to $S^{sys\prime}$ in any order. For this example, we will add the circulator first and then the attenuator, resulting in the sparse matrices shown in (29) and (30) below, but we could have entered them in the opposite order.

$$S^{sys\prime} = \begin{bmatrix} 0 & 0 & S_{13} & & \\ S_{12} & 0 & 0 & & \mathbf{0s} \\ 0 & S_{32} & 0 & & \\ & & & 0 & A_{12} \\ & \mathbf{0s} & & A_{21} & 0 \end{bmatrix} \quad (29)$$

$$C^{sys\prime} = \begin{bmatrix} T_{circ}^{cw}\begin{bmatrix} (1-|S_{13}|^2) & 0 & 0 \\ 0 & (1-|S_{21}|^2) & 0 \\ 0 & 0 & (1-|S_{32}|^2) \end{bmatrix} & \mathbf{0s} \\ \mathbf{0s} & T_{att}^{cw}\begin{bmatrix} (1-|A_{12}|^2) & 0 \\ 0 & (1-|A_{21}|^2) \end{bmatrix} \end{bmatrix} \quad (30)$$

The multiport representation of $S^{sys\prime}$ and $C^{sys\prime}$ are now complete, but the attenuator/circulator connection has not yet been defined. The next step is to define how the ports are to be connected to each other. Then, we must reorder matrix $S^{sys\prime}$ and $C^{sys\prime}$ according to how the ports are connected. We can choose how elements are connected by visualizing the connections we would like to take place within the multinetwork matrix $S^{sys\prime}$. Each row and column in $S^{sys\prime}$ represent a port in the multinetwork system shown in Fig. 6.

In this example, due to the order in which the elements are entered into matrix $S^{sys\prime}$, we want to connect the circulator (port 2) to one end of the attenuator (port 4).

The permutation matrix $P$ was intentionally ordered with external ports 1, 5, and 3. This is so that when $S^{net}$ and $C^{net}$ are calculated, port 5 in the multinetwork matrix $S^{sys\prime}$ will end up in the position corresponding to external port 2 in $S^{net}$. The external ports can be arbitrarily rearranged to the user's benefit using the permutation matrix $P$. In this example, we rearranged the ports to match the port convention outlined in Fig. 7. Matrices $K\prime$ and $P$ are defined for this example as:

$$K' = \begin{bmatrix} 0 & 0 & 0 & 0 & 0 \\ 0 & 0 & 0 & 1 & 0 \\ 0 & 0 & 0 & 0 & 0 \\ 0 & 1 & 0 & 0 & 0 \\ 0 & 0 & 0 & 0 & 0 \end{bmatrix} \tag{31}$$

$$P_e = \begin{bmatrix} 1 & 0 & 0 & 0 & 0 \\ 0 & 0 & 0 & 0 & 1 \\ 0 & 0 & 1 & 0 & 0 \end{bmatrix} \tag{32}$$

$$P_i = \begin{bmatrix} 0 & 1 & 0 & 0 & 0 \\ 0 & 0 & 0 & 1 & 0 \end{bmatrix} \tag{33}$$

$S_{ee}^{sys}, S_{ie}^{sys}, S_{ei}^{sys}, S_{ii}^{sys}, K_{ii},$ and $\Lambda$ can be calculated using (18) – (23). The solution of this set of equations yields:

$$S^{net} = \begin{bmatrix} 0 & 0 & S_{13} \\ A_{12}S_{21} & 0 & 0 \\ 0 & A_{12}S_{32} & 0 \end{bmatrix} \tag{34}$$

$$C^{net} = \begin{bmatrix} C_{11}^{net} & 0 & 0 \\ 0 & C_{22}^{net} & 0 \\ 0 & 0 & C_{33}^{net} \end{bmatrix} \tag{35}$$

where

$$C_{11}^{net} = k_B T_{circ}^{cw}(1 - |S_{13}|^2),$$

$$C_{22}^{net} = k_B T_{circ}^{cw}|A_{12}|^2(1 - |S_{21}|^2) + k_B T_{att}^{cw}(1 - |A_{12}|^2),$$

$$C_{33}^{net} = k_B T_{att}^{cw}|S_{32}|^2(1 - |A_{12}|^2) + k_B T_{circ}^{cw}(1 - |S_{32}|^2).$$

$S^{net}$ and $C^{net}$ ((34) and (35)) provide the scattering and noise parameters for the schematic described in Fig. 7. The temperatures $T_{circ}^{cw}$ and $T_{att}^{cw}$ are the noise temperatures defined in (24). Comparing (25) and Fig. 7 to (34), $S^{net}$ looks just like an $S_{circ}$ only with more attenuation associated with port 2 as expected. The path from port 3 to port 1 remains unchanged.

The next step is to verify that $C^{net}$ yields the correct noise covariance matrix. If correct, because the devices used in this example are ideal, the solution should be identical to the Friis formula for cascading noise temperature. To verify $C^{net}$, we can calculate the system noise temperature (in a 1 Hz bandwidth) for each port $T_i^{sys}$ [2]:

$$T_i^{sys} = \frac{\overline{cc^\dagger}_{ii}}{k_B S A S^\dagger_{ii}} = \frac{C^{net}_{ii}}{k_B S^{net} A S^{net\dagger}_{ii}} \tag{36}$$

The system noise temperature $T_i^{sys}$ is not related to the physical or noise temperature; rather, it is a way to express the level of available noise power introduced by the system being analyzed. The subscript $i$ in (36) denotes the port of interest in the system under analysis and $S$ is the s-parameters for the device. In our example, $S = S^{net}$ for the analysis in (35)–(37) below.

Calculating the noise temperature for each port we get:

$$T_1^{sys} = T_{circ}^{cw}\left(\frac{1}{|S_{13}|^2} - 1\right) = T_{circ\ 3\rightarrow 1}^{sys} \tag{37}$$

$$T_2^{sys} = T_{circ}^{cw}\left(\frac{1}{|S_{21}|^2} - 1\right) + \frac{T_{att}^{cw}\left(\frac{1}{|A_{12}|^2} - 1\right)}{|S_{21}|^2}$$
$$= T_{circ\ 1\rightarrow 2}^{sys} + \frac{T_{att}^{sys}}{C_{circ\ 1\rightarrow 2}} \tag{38}$$

$$T_3^{sys} = T_{circ}^{cw}\left(\frac{1}{|A_{12}|^2} - 1\right) + \frac{T_{att}^{cw}\left(\frac{1}{|S_{32}|^2} - 1\right)}{|A_{12}|^2}$$
$$= T_{att}^{sys} + \frac{T_{circ\ 2\rightarrow 3}^{sys}}{C_{att}} \tag{39}$$

There are three system noise temperatures that can be calculated, one for each port, whose paths are defined in Fig. 7 and are derived in (37)–(39).

For an ideal system, the noise temperature of a passive two-port device with a noise temperature $T_{device}^{CW}$ can be expressed in the form:

$$T_{device}^{sys} = T_{device}^{CW}(L - 1) \tag{40}$$

where loss is expressed as a positive number $L > 1$.

Using the paths traced out in Fig. 7, it can be demonstrated that the system noise temperature agrees with the Friis cascade for ideal two-port devices shown in Fig. 6 and (38) and (39).

$$T_{total}^{sys} = T_{device1}^{sys} + \frac{T_{device2}^{sys}}{G_{device1}} \tag{41}$$

This example was performed with perfectly ideal components, and so in this case a two-port Friis stack-up will also yield an identical solution of this example. When device isolation or return loss is nonideal, the overall system noise performance can be very different than the Friis cascade.

## VI. Model vs. Measurement Capturing the Thermal Transitions in a Dilution Refrigerator

Simulated data are used to predict the measurements outlined in this section. For simulation, scikit-RF [10]—an open-source Python package for RF and Microwave applications—was used. In particular, the technique detailed in this paper has been incorporated into a forked branch of scikit-rf [11]. Although many methods exist [9], [12]–[15], the approach detailed in [2] was used to derive noise parameters. The experiment is set up to illustrate this method for relevant cryogenic research. Leading-edge research related to superconducting qubits [16], microwave quantum sensing [17], and wave-like dark matter detection [18]–[20], for example, relies on the cooling power of a dilution refrigerator to perform sensitive, low-temperature radio frequency (RF) measurements.

However, unlike connectorized room-temperature RF systems that can be easily tested in pieces and then assembled for further testing, cryogenic systems are difficult to dissect. There are several issues that make cryogenic RF measurements challenging. *First, components cannot easily be tested in isolation when cold, leading to de-embedding challenges.* Cables, bulkhead feedthroughs, attenuators, and sometimes more complicated devices are integral to the measurement, often for shielding reasons. Care must be taken to heavily filter, thermally sink, and attenuate twisted pair and RF coax wiring at each stage to prevent raising the base temperature of the refrigerator which has limited cooling power at the bottom stage. In terms of RF signals, typically signals injected into the refrigerator are attenuated at various thermal stages (totaling 40–80 dB) to thermalize the coax lines and prevent the mixing chamber stage from being exposed to 290 K (room temperature) blackbody radiation. RF output signals, which cannot be attenuated for signal-to-noise reasons, are typically both amplified at low-temperature stages and sent through well-thermalized isolators and 0 dB attenuators to prevent 290 K heating propagating backward down the output. *Second, components behave differently when cold.* Cables are less lossy, conductors shrink, circulator band-pass frequencies shift, amplifier biases change, and certain components like superconducting parametric amplifiers do not operate at room temperature and thus must be measured at cryogenic temperature. *Third, devices within the cryogenic space are often at very different temperature stages, leading to a complicated cascade of noise sources.* Issues like impedance mismatch and device isolation parameters can make noise analysis difficult across thermal barriers.

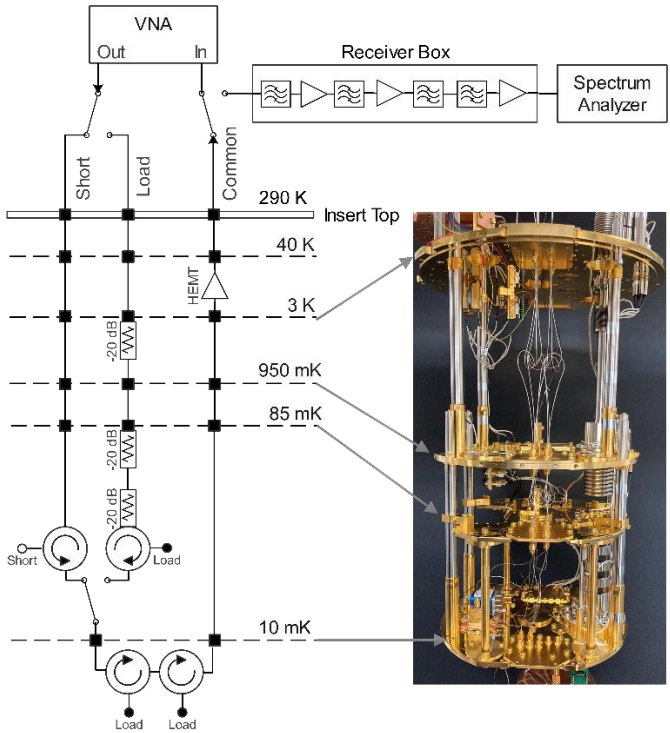

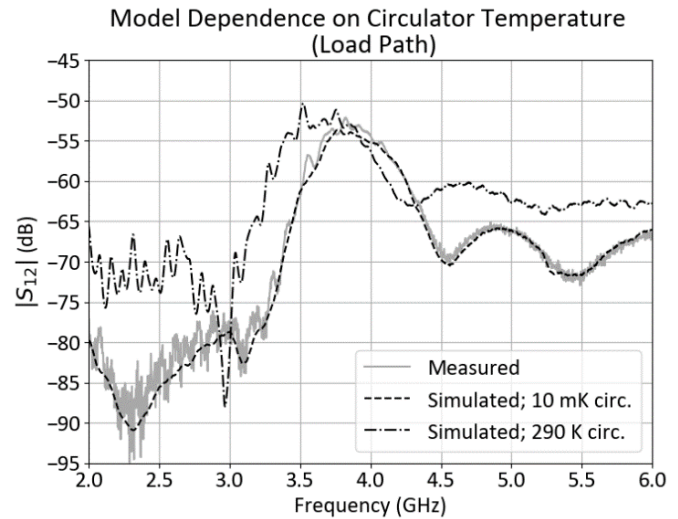

Fig. 9. Insertion loss ($|S_{12}|$) as a function of frequency for the load path. Measured data (gray solid line), simulated result with 10 mK circulator data (black dashed line), and simulated result with 290 K circulator data (black dash-dot line) are shown. Model dependence on circulator temperature can be seen. Simulated result with 10 mK circulator data exhibits better agreement with the measured data. Simulated result with 290 K circulator data appears to have about 300 MHz shifted compared to the measured data.

Fig. 8. Schematic of demo cryogenic experiment. Three 20 dB attenuators (Bluefors CRYO attenuator) are shown in the load path at 10 mK, 85 mK, 3 K. The cryogenic high-electron-mobility transistor (HEMT) amplifier (LNF-LNC_3_14A [21]) is at 3 K in the common path. Dashed lines represent common thermal stages. Black blocks represent SMA connection on each thermal stage (bulkhead on flanges). Receiver box contains room-temperature electronics, four filters (MC 15542 VBFZ-4000-s+ [22]), and three amplifiers (ZX60-83LN-S+ [23]). Gray arrows are connecting thermal stages in the schematic to corresponding parts in the insert picture (right).

These challenges complicate the measurement process and motivate our need for computer modeling tools for cryogenic

RF systems. We claim that the algorithm detailed in this work is relevant to the above challenges. We devised the experiment illustrated in Fig. 8 to demonstrate the capabilities of this method in a cryogenic system.

In this experiment, designed to mimic the typical supporting infrastructure needed to operate a superconducting device, our objective was to perform a difference measurement (similar to a Y-factor measurement [24]), where a shared amplification stage output (termed the "common" path) would be switched between two nearly identical but intentionally distinct paths. These two lines, the "short" side and the "load" side, are named after a key asymmetry on the second port of their input circulators: a short and a 50 Ω load.

The only other difference between the paths is 60 dB of attenuation on the load side distributed across the different thermal stages (20 dB at 10 mK, 85 mK, and 3 K following typical thermal sinking practices) and no attenuation on the short side (breaking convention). Standard cryogenic practices recommend placing attenuators on each thermal stage to help thermally tie the center conductor to each temperature stage. Here, all attenuation was removed on the "short" path to intentionally exaggerate black-body effects to contrast with the "load" path.

These two key asymmetries were implemented to create a hot/cold system temperature difference that can be seen by a spectrum analyzer on the amplified output of the common path. When switched to the "hot" short side, the system temperature will be affected by room temperature radiation coming down the coax, interacting with the passive devices at different temperatures, reflecting off the shorted port and up the common path. By contrast, the "cold" load side will see a much lower system temperature; it should be dominated by 10 mK thermal noise at the mixing chamber stage. The common path between the two switch ports consisted of two isolators—added for thermal isolation and preemptively in anticipation of future work using sensitive superconducting RF devices—and a high-electron-mobility transistor (HEMT) amplifier at 4 K.

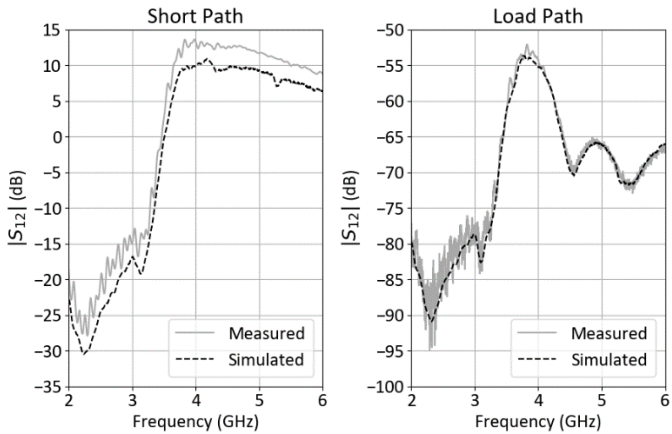

Fig. 10. Insertion loss ($|S_{12}|$) as a function of frequency plots. Simulated results and measured values are compared for short path (left) and load path (right). Measured data are in gray solid lines and simulated results are in black dashed lines.

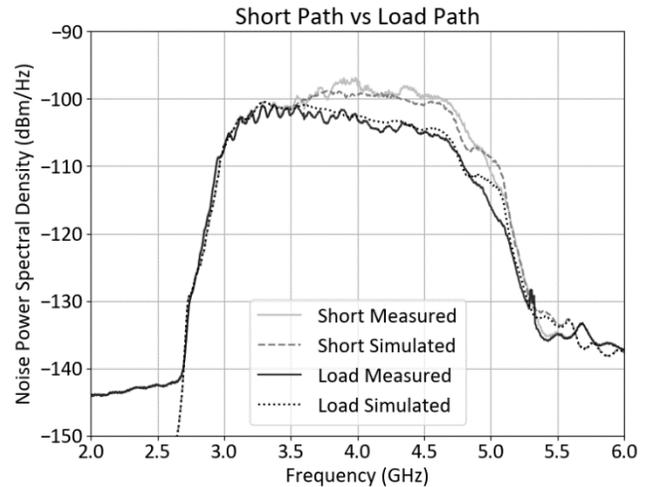

Fig. 11. Power spectral density as a function of frequency as seen at the input to the spectrum analyzer. Both short path (measured in gray solid line and simulated in gray dashed line) and load path (measured in black solid line and simulated in black dotted line) are presented.

For the cascaded model, each component was measured and cascaded using the method outlined in this paper. All passive items (cables, bulkheads, filters, attenuators, etc.) were measured at room temperature. For cables at temperatures less than 4 K, we used information provided by Bluefors on cable attenuation at cryogenic temperatures as the loss model [25]. For the HEMT amplifier, we used the vendor-supplied data along with method defined in [2] to simulate the noise covariance diagonals for all amplifier ports.

The circulator model was originally constructed using room-temperature data. However, a discrepancy between the full cascaded model and vector network analyzer (VNA) s-parameter measurements of the experiment revealed a temperature dependent 270 MHz shift caused by the circulators. To account for this, in a second cryogenic run, circulators were measured at 10 mK and an assumption about cable path symmetry was used to de-embed the circulators (removing cable effects) to reconstruct a circulator model at base temperature. Fig. 9 illustrates the impact that circulator temperature has on the response of the entire load side.

A completed cascaded model for both the short and load paths was compared to VNA measurements where the experiment was toggled between the two paths using a cryogenic Radiall switch (R577.432.000). Good agreement between measured and simulated results can be seen in Fig. 10. Both paths are marked by a highly attenuated left shoulder where circulators fall out of spec below 4 GHz. The short path has an anticipated highly conductive stack-up from the lack of attenuation while the load side exhibits ripples from the terminated port.

Simulated and measured results are within 2–5 dB as seen in Fig. 10; these errors could be from many possible sources. One such source is the measurement of the circulators in the dilution fridge. For the 10 mK circulator data, a series of cables were cascaded together to take data with room-temperature equipment. These cables were subtracted using an s-parameter de-embedding technique. A cascaded set of RF cables of equal length were looped in through the dilution fridge and measured separately as a control. Each circulator in the system was measured separately with a supposedly equal length of cables and then the control cable data were used as the through-measurement for de-embedding. The technique assumes that these de-embedding cables are identical in return and insertion losses to those used in the circulator measurements. This assumption has the potential to yield errors. It is possible that more detrimental loss errors were introduced when de-embedding the short path than the load path. Other possible sources of error include unaccounted for adaptors or losses between measuring individual components and the overall system.

For noise measurements, an extra stage of amplification, labeled "Receiver Box" on the top of Fig. 8 was added, so the measured thermal noise in the dilution refrigerator would be above the spectrum analyzer noise floor. For these external amplifiers, we measured the s-parameters using a VNA along with noise figure data supplied by the vendor to determine the noise covariance matrix following [2]. We used filters to prevent the wideband amplifiers from saturating.

We simulated the expected power spectral density at the input to the spectrum analyzer using the method outlined in this paper and compared to the data measured by the spectrum analyzer in units of power spectral density (Fig. 11). The spectrum analyzer data were divided by the resolution bandwidth used for the measurement of the data (1.1 MHz) to arrive at a power spectral density for comparison. Measurement and simulation results show good agreement, even in the out-of-band upper tail. This demonstration shows that power spectral density shape can be predicted by this temperature-dependent method. The noise measurement does not show the same sensitivity as the error in s-parameters but instead is more sensitive to thermal variations.

Several assumptions about the cables were made to calculate the noise power spectral density. Unfortunately, all cables were measured at room temperature. The cable losses at or less than 4 K were reduced by about a factor of two to account for the cryogenic behavior of cables. Also, the physical temperature of the cables was speculated. This was difficult considering that the cables were thermally connected to two different thermal plates. The noise in Fig. 11 can vary considerably based on the assumed temperature of these cryogenic cables. Though the general outline of the data is consistent, the overall offset in power spectral density can be affected.

Originally it was assumed that the cable physical temperature was the mean between the two thermal plates it was attached to, but this approach yielded a power spectral density much higher than was measured. The plot in Fig. 11 is the result of assuming that each cable's physical temperature is equal to the lowest temperature thermal plate it was attached to. More investigation is needed to provide better physical cable temperatures. It is worth noting that all other components were well-thermalized and mounted to thermal plates. The temperature sensors of the respective plates were used in their noise covariance calculations.

## VII.   FUTURE WORK

The measurements were designed to verify performance of the algorithm at a preliminary level. A detailed analysis—proper error and uncertainty analysis like [26]—is left for future work. The simulation performed in the previous section was mostly a combination of room-temperature measurements and information from vendor data sheets. Only the circulators were measured at 10 mK. The example illustrated that with proper data one can closely approximate the behavior of a complex multiport system that extends over many operating temperatures.

Some of the noted sensitivities and errors in the measured versus simulated results suggest the need for better RF measurement methods for cryogenic components as well as improved understanding of physical temperature gradients along RF cables and the relationship of that gradient to the noise temperature.

Future simulation work will focus on developing, improving, and verifying simulated models for complex active and passive devices operating at cryogenic temperatures. This includes modeling superconducting cables and circulators, and new devices such as Josephson parametric amplifiers. Having better simulations of these types of components will aid in better system-level engineering of future experiments and reduce the need to rely on measured data.

## VIII.   CONCLUSION

In this paper we proposed using a noise-wave cascade method for the simulation of noise parameters and s-parameters of complex multiport cryogenic systems. Though other methods have also been proposed [27], this method seems to be unique in its approach and simplicity. We outlined the proposed method and presented the algorithm in step-by-step detail so that the process can be easily coded into an RF CAD program for use in simulation and modeling.

Measurement and simulation agreed within a reasonable tolerance for use in RF systems modeling. The measurement also illustrated the potential use of simulation in RF systems that include cryogenic components.

This paper also outlined future work to improve the fidelity of the modeling as well as the possibility of including future ideal quantum component models into future noise and s-parameter simulations.

ACKNOWLEDGMENT

The authors are grateful to M. Jones, Dr. VanDevender, Dr. Bunker, and Dr. Smith for providing advice and encouragement. This research benefited from use of a PNNL dilution refrigerator, operated under the Chemical Dynamics Initiative. The authors also would like to thank Dr. Direen for helping to seamlessly integrate the algorithm into the scikit-rf environment (scikit-rf.org) through a forked branch, allowing for its ease of use in RF simulations used to generate the plots in this paper.